# Eigenmode Decomposition Method for Full-Wave Modeling of Microring Resonators

Yuriy Akimov[1,*], Aswin Alexander Eapen[2], Shiyang Zhu[3], Doris K. T. Ng[3], Nanxi Li[3], Woon Leng Loh[3], Lennon Y. T. Lee[3], Alagappan Gandhi[1], and Aravind P. Anthur[2]

[1]*Institute of High Performance Computing (IHPC), Agency for Science, Technology and Research (A*STAR), 1 Fusionopolis Way, #16-16 Connexis, Singapore 138632.*
[2]*Institute of Materials Research and Engineering (IMRE), Agency for Science, Technology and Research (A*STAR), 2 Fusionopolis Way, #08-03 Innovis, Singapore 138634.*
[3]*Institute of Microelectronics (IME), Agency for Science, Technology and Research (A*STAR), 2 Fusionopolis Way, Innovis #08-02, Singapore 138634.*
[*]akimov@ihpc.a-star.edu.sg

We develop a theoretical predictive model for an all-pass ring resonator that enables the most complete description of linear coupling regimes. The model is based on eigenmode decomposition of Maxwell's equations with full account of the confined and leaky modes, as opposed to the existing phenomenological methods restricted to the confined modes only. This model enables quantitative description of all-pass ring resonators and provides insights into the physics underlying microring-waveguide coupling. We experimentally validate the model using transmission measurements in the linear regime of aluminium nitride resonators. The developed model is then used to explore the field enhancement in microrings crucial for nonlinear photonic applications.

## I. INTRODUCTION

Microring resonators are small ring-shaped waveguiding structures that can trap and manipulate light. Their compact size, low power consumption, ease of fabrication, and high-speed operation make them ideal for use in integrated photonic circuits, enabling the development of next-generation communication systems, sensors, biomedical devices, and quantum technologies [1]. Microring resonators play a crucial role in nonlinear photonics [2]. They can resonantly enhance nonlinear optical processes giving rise to new effects not possible with conventional waveguides, such as frequency comb generation [3-4]. Microring resonators are also essential in nonlinear quantum photonics [5], where they are used to generate and manipulate non-classical light states [6], such as entangled photon pairs and squeezed light [7]. They can also create nonlinear interactions between single photons, which are important for quantum logic gates and quantum simulations [8]. The ability of microring resonators to enhance nonlinear effects and confine light in a small space makes them an important tool for advancing research in quantum technologies and developing practical applications. As researchers continue to explore their potential in new and exciting applications, the importance of microring resonators in photonics and quantum optics is expected to further grow.

Although resonant optical effects in microrings have been extensively reported experimentally, theoretical studies of them are comparatively scarce due to the complexity of their description. There are three groups of phenomenological methods developed for linear and nonlinear regimes of microring resonators: (i) Maxwell's-equations-based methods [9,10], (ii) Schrödinger-equation-based methods [11-13] and (iii) multiple port network methods [14-16]. The Maxwell's-equations-based methods benefit from their full-wave description but lack simplicity especially for nonlinear operation of microring resonators. The Schrödinger-equation-based methods benefit from simpler treatment provided for slowly varying envelopes and adapted for nonlinear studies for the price of lower accuracy. The multiple port network methods enable analytical description of the linear regime of coupling but suffer from multiple severe simplifications.

Being approximate, all the methods developed have their own limitations caused by their natural inconsistencies. One of the inconsistencies common to the methods given in Refs. [9-16] is the single-point treatment of coupling. In the context of all-pass ring resonators, the single-point coupling means that the bus-waveguide and the microring interact with each other strictly on the cross-sectional plane passing through the position of minimal waveguide-ring gap. Before and after the coupling plane, the electromagnetic fields of the waveguide and the microring are assumed not to intersect. There were multiple attempts to improve the description by substituting the single-point coupling coefficients with those computed over extended coupling areas (e.g., see Refs. [16-18]). They helped in some aspects but left the core physics of the methods unchanged with the same internal inconsistencies.

Another common issue is the forward modeling of signal propagation. The methods given in Refs. [9-16] ignore feedback from the out-coupled fields to the ring resonator. In this paper, we will demonstrate that the feedback of the out-coupled fields is crucial for the ring mode dynamics, defining the gap-size dependence of the microring resonances.

The last issue to point out is the incomplete mathematical treatment provided for the microring fields in methods [9-16]. Azimuthal wavenumbers of those fields appear continuous, violating the periodicity condition for the supported ring fields and making them mathematically inconsistent.

Altogether, the mentioned inconsistencies make the methods developed suitable mainly for fitting of experimental data, rather than for predictive modeling and design of optical resonators. To make predictive modeling possible, full-wave solution of Maxwell's equations for entire all-pass microring resonator is required. Though such a solution can potentially be obtained numerically [17-20], huge size of typical resonators makes the full-wave simulation computationally expensive if not barely possible. This situation urges development of analytical or semi-analytical methods allowing quantitative predictions without computationally expensive simulations.

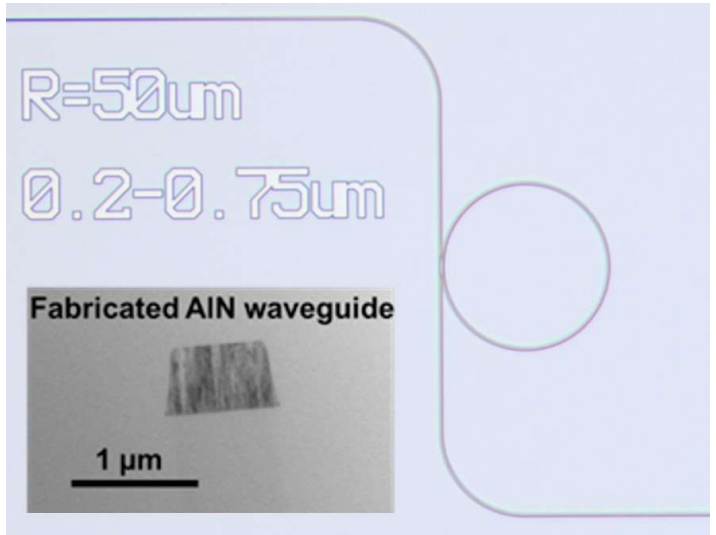

**Fig. 1.** An optical microscope image of the fabricated AlN resonator with the radius of 50 μm. The inset shows cross-sectional TEM image of the fabricated AlN waveguide.

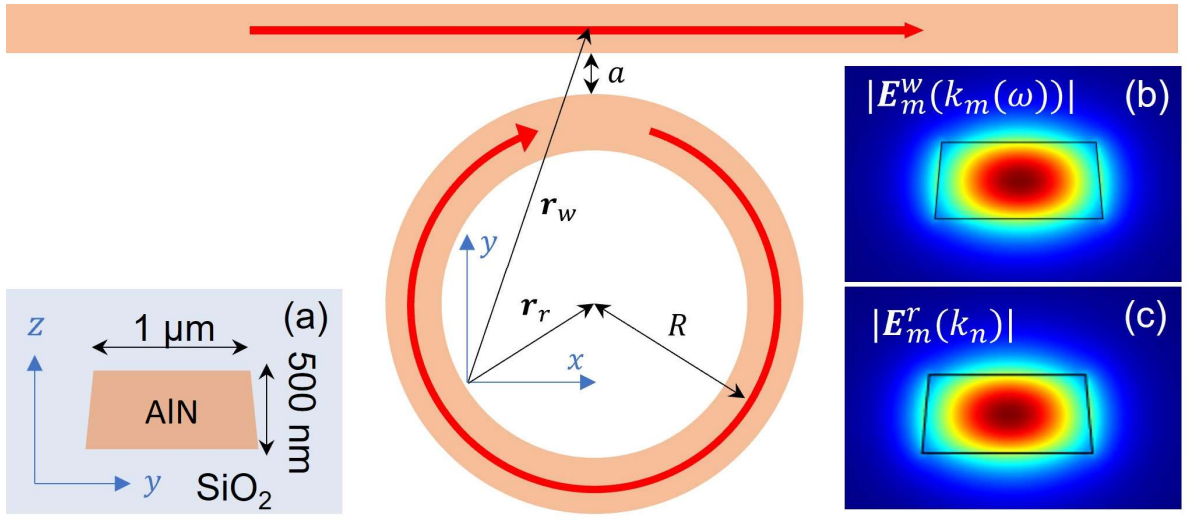

**Fig. 2.** Sketch of the model system marked with the notations used. Insets give the cross-sectional dimensions (a) and the field distribution in the waveguide (b) and the ring (c).

As opposed to phenomenological approaches [9-16], we develop a systematic, fully Maxwell's-equations-based predictive model for all-pass ring resonators. The model provides a complete theoretical representation of the microring resonator and delivers an accurate linear theory in both overcoupling and undercoupling regimes. We validate the model by comparing its results with experimental measurements of the transmittance quality factors obtained in the linear regime of aluminium nitride (AlN) microresonators around the 1.3 μm wavelength. The comparison demonstrates excellent agreement between the theoretical and experimental results. By using the developed model, we analyze and optimize the performance of AlN ring resonators for maximum fields excited inside the microrings. The predicted optima are further compared with the experimental data and confirmed by the observed extinction maxima.

An optical microscope image of the fabricated AlN microresonator with the radius of 50 μm is shown in Fig. 1. The inset shows the cross-sectional transmission electron microscopy (TEM) image of the fabricated AlN waveguide cladded in $SiO_2$ (see Appendix G for fabrication details). All the resonators are end-fire coupled, with light injected using a lensed fiber. AlN is chosen as the resonator material for its wideband operation from ultra-violet to infrared wavelengths [21], large bandgap (~6.2 eV) [22] and strong second order nonlinearity ($\chi^{(2)} \sim 4.7$ pm/V) [23]. On top of these, AlN thin films can be scalable up to 8-inch or 12-inch diameters depending on the wafer size that it is deposited on, increasing device output yield of integrated photonics and ring resonators in one fabrication round. Also, AlN benefits from low temperatures used for its deposition that can be as low as 200°C [24].

## II. WAVEGUIDE-RING COUPLING MODEL

A described all-pass ring resonator is composed of a bus straight waveguide and a microring of the same cross-sectional profile and the same material composition. The device dimensions and the field profiles used in experimental validation of the model are shown in Fig. 2.

The developed model is based on full-wave solution of the linearized Maxwell's equations (see Appendix A for details). For description of the ring-waveguide coupling, we ignore all intrinsic dissipation in the structure and decompose the fields excited inside the ring and the waveguide over their eigenfields $\boldsymbol{E}_m^r(\boldsymbol{r},k_n)$ and $\boldsymbol{E}_m^w(\boldsymbol{r},k)$:

$$\boldsymbol{E}_r(\boldsymbol{r},\omega) = \sum_n \sum_m A_m^r(k_n,\omega)\boldsymbol{E}_m^r(\boldsymbol{r},k_n), \tag{1}$$

$$\boldsymbol{E}_w(\boldsymbol{r},\omega) = \int_{-\infty}^{\infty} dk \sum_m A_m^w(k,\omega)\boldsymbol{E}_m^w(\boldsymbol{r},k), \tag{2}$$

where $A_m^r(\omega,k_n)$ and $A_m^w(\omega,k)$ are the frequency- and wavenumber-dependent complex mode amplitudes of the microring and the bus waveguide. In Eqs. (1) and (2), $\boldsymbol{E}_m^r(\boldsymbol{r},k_n)$ and $\boldsymbol{E}_m^w(\boldsymbol{r},k)$ are the eigen solutions of Maxwell's equations for the single microring and waveguide respectively. The eigenfields of the microring and the waveguide oscillate at the real-valued eigenfrequencies given by $\omega_m(k_n)$ and $\omega_m(k)$, where $m$ is the integer modal index, $k_n = n/R$ is the discrete azimuthal ring wavenumber, and $k$ is the continuously varying longitudinal waveguide wavenumber. Details of the field decomposition over the waveguide and ring eigenfields are given in Appendices B and C.

The real-valued eigenfrequencies make the respective eigenfields orthogonal in the Hilbert space enabling a fast and easy way for finding the mode amplitudes $A_m^w(\omega,k)$ and $A_m^r(\omega,k_n)$. Using the eigenfield orthogonality, the dissipation-free Maxwell's equations can be reduced to the following master equations for the coupled dynamics of the mode amplitudes:

$$\left[\frac{\omega_m^2(k_n)}{\omega^2} - 1\right] A_m^{w,in}(\omega,k) = A_m^0(\omega,k), \tag{3}$$

$$\left[\frac{\omega_m^2(k_n)}{\omega^2} - 1\right] A_m^{w,out}(\omega,k) = R\sum_{n=-\infty}^{\infty} \theta_{mm}^{out}(k_n,k)A_m^r(\omega,k_n), \tag{4}$$

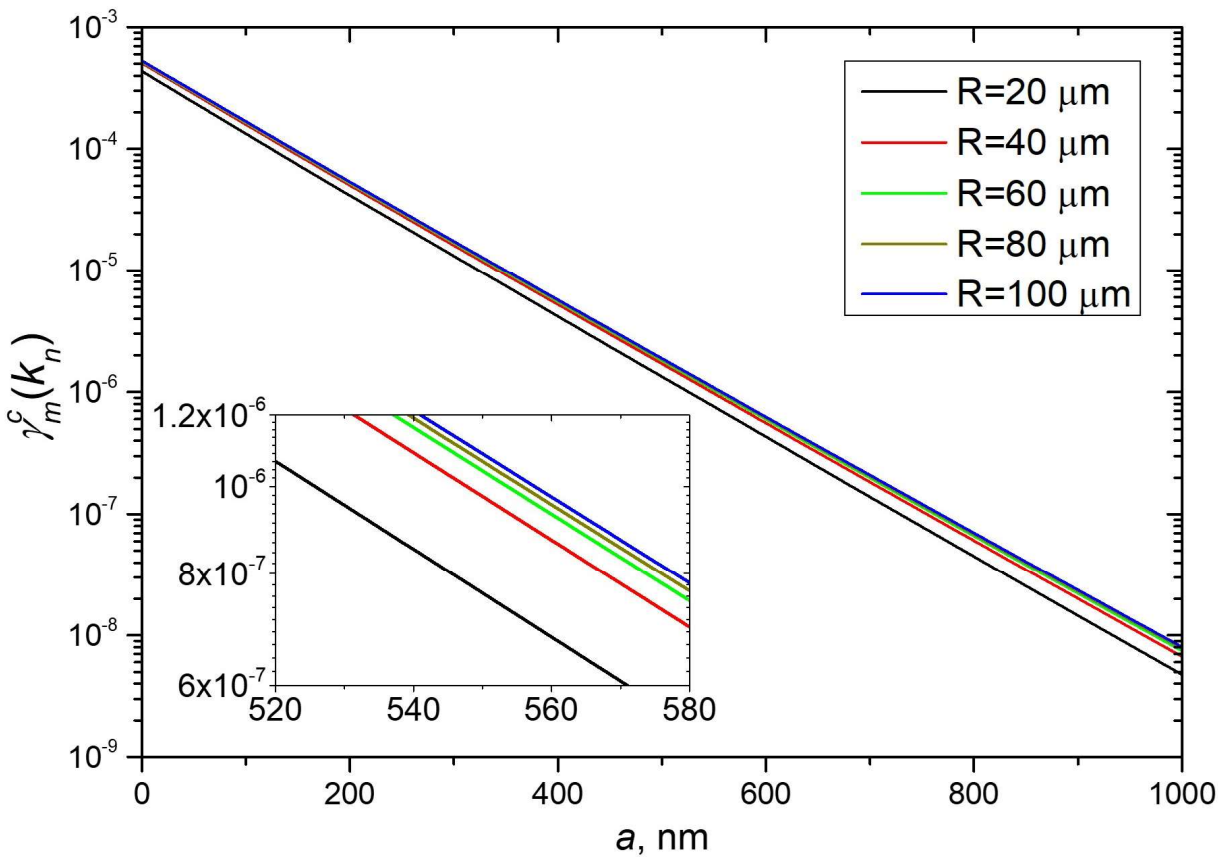

**Fig. 3.** The coupling loss rates of the microrings eigenmodes around the 1.3 μm wavelength predicted by the theoretical model.

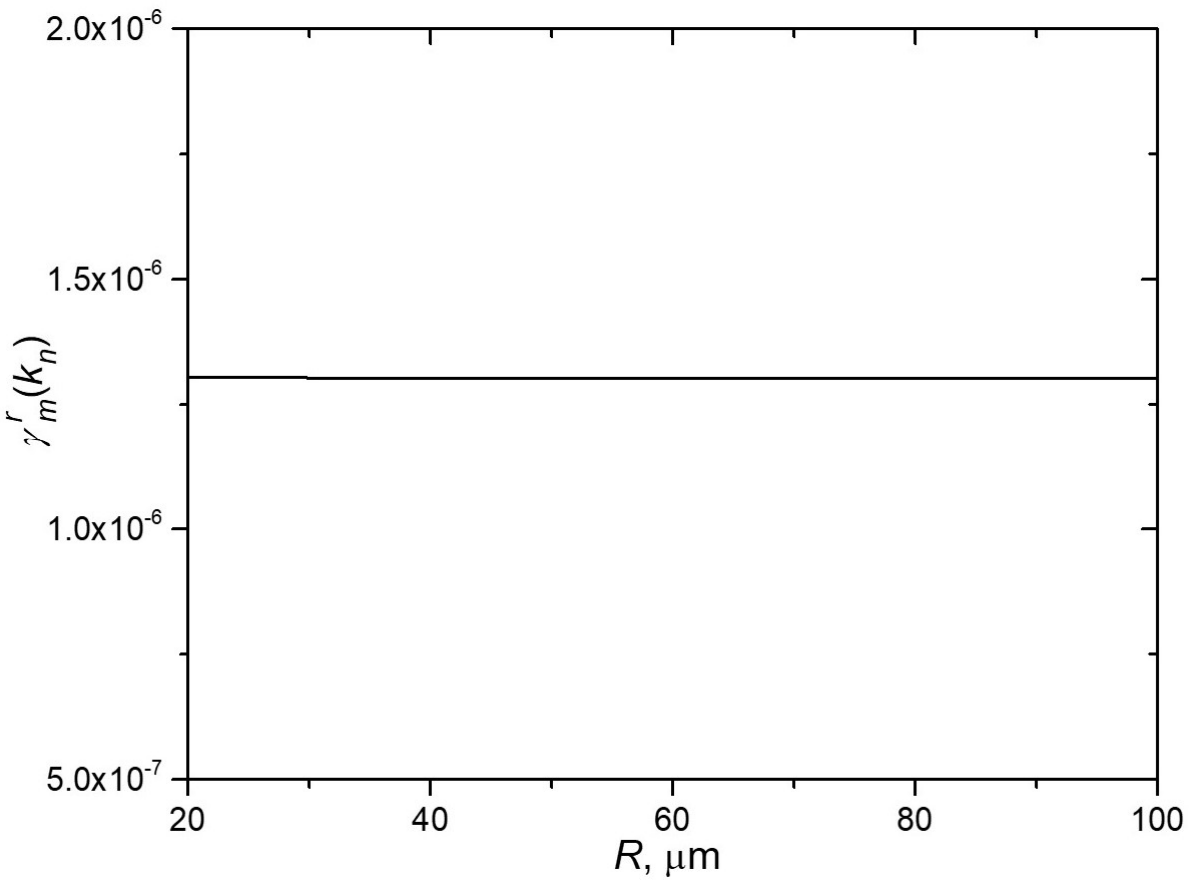

**Fig. 4.** The intrinsic loss rates of the microrings eigenmodes around the 1.3 μm wavelength obtained in numerical modeling.

$$\left[\frac{\omega_m^2(k)}{\omega^2} - 1\right] A_m^r(\omega, k_n) = \int_{-\infty}^{\infty} dk\, \theta_{mm}^{in}(k_n, k) A_m^w(\omega, k), \quad (5)$$

where the fields excited in the microring and the bus waveguide are assumed to be given by the modes with the same index $m$. In these equations, we intentionally split the amplitudes $A_m^w(\omega, k) = A_m^{w,in}(\omega, k) + A_m^{w,out}(\omega, k)$ into the two parts $A_m^{w,in}(\omega, k)$ and $A_m^{w,out}(\omega, k)$ to separate the effects of (i) the external source $A_m^0(\omega, k)$ pumping the waveguide and (ii) the out-coupling of the ring fields $A_m^r(\omega, k_n)$ that similarly pump the waveguide fields. As for the ring modes fields, they are pumped by the waveguide modes with amplitudes $A_m^w(\omega, k)$. The processes of in-coupling and out-coupling between the ring and waveguide modes are given with the complex coefficients $\theta_{mm}^{in}(k_n, k)$ and $\theta_{mm}^{out}(k_n, k)$. Calculation of all these parameters is given in Appendix D.

Noteworthy that master equations (3)-(5) provide mathematically complete description of energy exchange between the discrete microring modes and the continuously distributed over $k$ waveguide modes. For a fixed frequency $\omega$, only the modes with $k = k_m(\omega)$ [being the inverse function of $\omega = \omega_m(k)$] are eigen and confined inside the waveguide and the ring. Modes with wavenumbers other than $k_m(\omega)$ are generally leaky bringing coupling-induced radiation losses to the fields excited in the bus waveguide and the ring resonator. Inclusion of both the confined and leaky harmonics in the field decomposition makes the master equations the most complete field description for the waveguide-ring coupling compared to the existing phenomenological methods that consider the confined modes only.

## III. RING MODE LOSSES

### A. Coupling losses

Under weak coupling, when the in-coupling and out-coupling coefficients are sufficiently small: $|\theta_{mm}^{in}(k_n, k)|$, $|\theta_{mm}^{out}(k_n, k)| \ll 1$, the master equations predict resonant excitation of the ring modes. The ring mode with the azimuthal wavenumber $k_n$ experiences the resonance around the frequency $\omega = \omega_m^{res}(k_n) \approx \omega_m(k_n)$ with the $Q$-factor of

$$Q_m^c(k_n) = [2\gamma_m^c(k_n)]^{-1}, \quad (6)$$

where $\gamma_m^c(k_n)$ is the coupling loss rate of the ring mode:

$$\gamma_m^c(k_n) = \frac{\pi R}{4}\left|\frac{\omega_m(k_n)}{V_m^g(k_n)}\right| \mathrm{Re}\,[\theta_{mm}^{in}(k_n, k_n)\theta_{mm}^{out}(k_n, k_n)], \quad (7)$$

with $V_m^g(k) = \partial\omega_m(k)/\partial k$ being the group velocity of the eigenwave with mode index $m$.

The coupling loss is the wave-interaction-induced attenuation mechanism of the ring modes. It originates from the coupled dynamics of the ring and waveguide modes. From the one side, a particular ring mode out-couples to all the confined and leaky waveguide modes. From the other side, the same waveguide modes in-couple to the considered ring mode back. The combined effect of the in-coupling and out-coupling results in an effective loss for every ring mode given by the integral contributions of the two processes and involving all the leaky and confined modes of the waveguide. Detailed derivation of the coupling loss rate $\gamma_m^c(k_n)$ is shown in Appendix E.

Being defined with the geometry-dependent in-coupling $\theta_{mm}^{in}(k_n, k_n)$ and out-coupling $\theta_{mm}^{in}(k_n, k_n)$ coefficients, the coupling loss rate $\gamma_m^c(k_n)$ varies with the waveguide-microring gap $a$ and the ring radius $R$, as shown in Fig. 3.

### B. Intrinsic losses

In addition to the coupling loss, all-pass ring resonators exhibit intrinsic losses of absorption, bending, volume and surface scattering for the ring and the waveguide. To account for these losses, we add the intrinsic decay rates $\gamma_m^r(k_n)$ and $\gamma_m^w(k)$ to the ring and waveguide eigenfrequencies $\omega_m(k_n) \to \omega_m(k_n)[1 - i\gamma_m^r(k_n)]$ and $\omega_m(k) \to \omega_m(k)[1 - i\gamma_m^w(k)]$. With intrinsic loss inclusion, eigenmode decompositions (1) and (2) get mathematically incomplete, so the coupling loss rate given by Eq. (7) holds as long as $\gamma_m^r(k_n), \gamma_m^w(k) \ll 1$. In

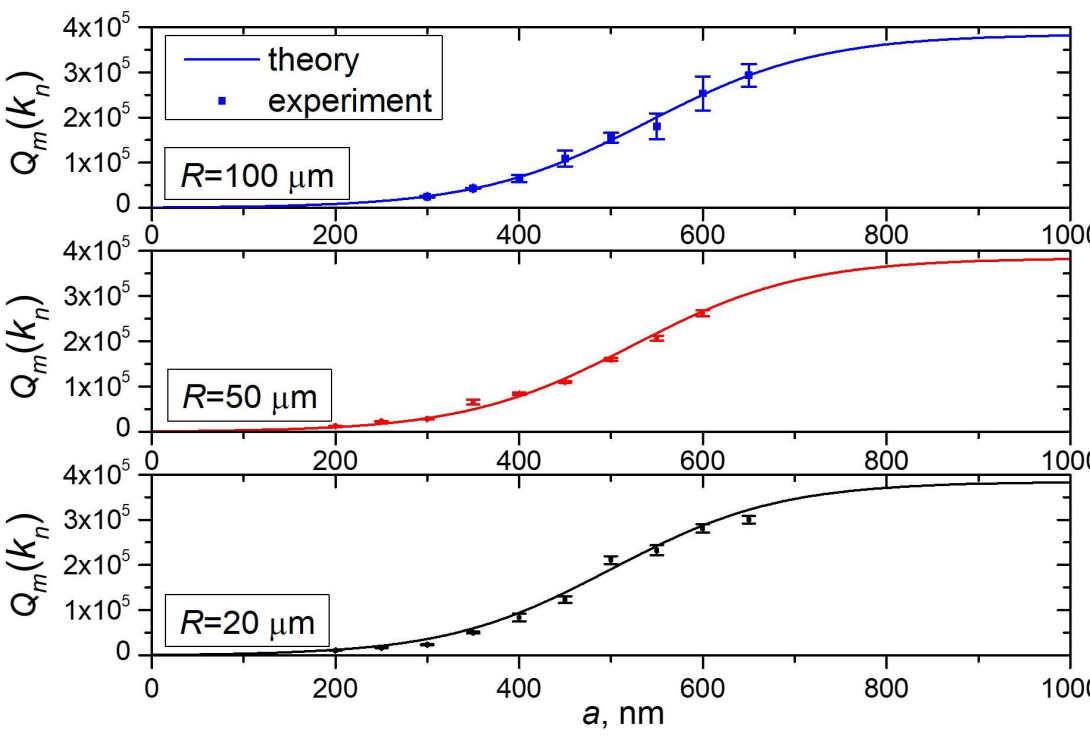

**Fig. 5.** Theoretical and experimental quality factors of optical resonances around the 1.3 μm wavelength in the AlN resonators with different radii and waveguide-ring gaps.

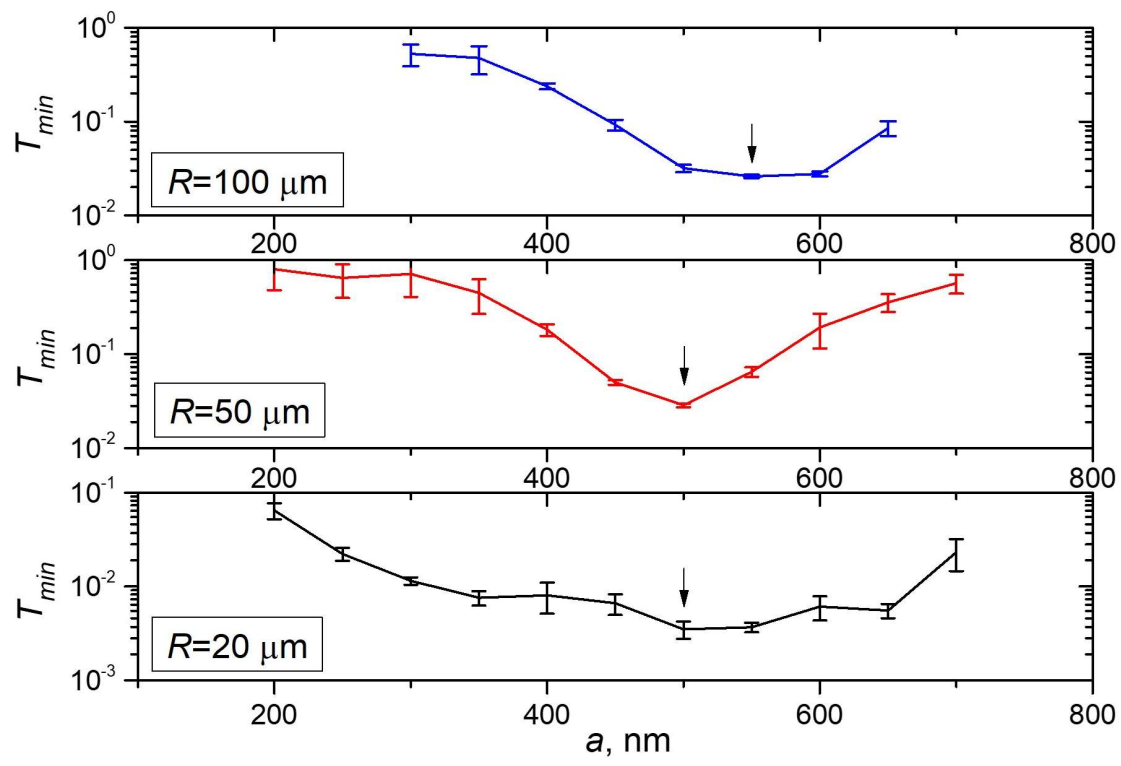

**Fig. 7.** Experimentally measured waveguide transmittance minima around the 1.3 μm wavelength as a function of ring-waveguide gap for different microring radii. The arrows show the minima with the highest microring-induced extinction.

this case, the intrinsic losses just contribute an additional $Q$-factor:

$$Q_m^r(k_n) = [2\gamma_m^r(k_n)]^{-1} \tag{8}$$

to the microring resonances.

To get the intrinsic loss rates of AlN microrings, we run COMSOL Multiphysics eigenmode solver for standing alone rings with the extinction coefficients of the materials used that reproduce the experimentally measured propagation loss of the waveguide modes, as discussed in Appendix I. The obtained rates $\gamma_m^r(k_n)$ are shown in Fig. 4. They account for the absorption, scattering and bending losses of the ring modes. The loss rates $\gamma_m^r(k_n)$ appear weakly dependent on ring size for $R \geq 20$ μm. The smallness of $\gamma_m^r(k_n)$ in the fabricated AlN microrings validates the use of Eq. (7) for estimation of the coupling loss rate $\gamma_m^c(k_n)$. By comparing $\gamma_m^r(k_n)$ and $\gamma_m^c(k_n)$, we expect the coupling loss to dominate for small waveguide-ring gaps with $a < 500$ nm and the intrinsic losses to prevail for large gaps with $a > 500$ nm.

### C. Total Losses

With account of the coupling and intrinsic losses, the total $Q$-factor of the microring resonances becomes

$$Q_m(k_n) = [2\gamma_m^c(k_n) + 2\gamma_m^r(k_n)]^{-1}. \tag{9}$$

Emphasize that among $\gamma_m^c(k_n)$ and $\gamma_m^r(k_n)$ only the coupling loss rate depends on the waveguide-microring gap $a$ and brings this dependence to $Q_m(k_n)$. As $\gamma_m^c(k_n)$ decreases with $a$ to zero, the highest quality factor $Q_m^{max}(k_n) = Q_m^r(k_n)$ is reached in the uncoupled regime under infinite gap $a$ when both the in-coupling and out-coupling naturally vanish.

To verify the model for prediction of total losses, we calculated $Q_m(k_n)$ as a function of gap $a$ and compared them with the quality factors experimentally measured around the wavelength of 1.3 μm for AlN all-pass ring resonators with the radii of 20, 50 and 100 μm radius (see Appendices F and H for details). The $Q$-factor comparison is shown in Fig. 5. It demonstrates an excellent agreement between the theoretical and experimental results for all the microring radii considered and, thus, validates our theoretical model.

## IV. FIELD ENHANCEMENT

Optical resonances are accompanied by generation of strong microring fields. This effect is used in nonlinear photonic applications to enhance efficiency of nonlinear processes. For narrowband profiles $A_m^{w,in}(\omega,k)$, the ring fields can be reduced to:

$$|A_m^r(\omega_m^{res}(k_n),k_n)| \approx \eta_m(k_n)\left|\int_{-\infty}^{\infty} dk\, A_m^{w,in}(\omega,k)\right|, \quad (10)$$

where

$$\eta_m(k_n) = Q_m(k_n)\left|\theta_{mm}^{in}(k_n,k_n)\right| \quad (11)$$

is the integral enhancement factor of the resonant ring mode. To get the highest fields, the integral enhancement should be maximized. The critical coupling that corresponds to the maximum fields excited occurs, when all incoming power goes into the resonant ring mode causing a drop in the bus waveguide transmission to zero (see Appendix F for details). Following our model, the highest fields generated in AlN microrings are achieved at $a_{cr}$ varying from 507 nm for $R = 20$ µm to 534 nm for $R = 100$ µm, as shown in Fig. 6. The maximum field enhancement $\eta_m^{cr}(k_n)$ observed is about 20 for $R = 20$ µm and drops to about 8 for $R = 100$ µm. The predicted values for $a_{cr}$ are in very good agreement with the experimentally measured bus waveguide transmittance minima shown in Fig. 7, following which the maximum microring-induced extinction is observed for the gaps in the range from 500 to 550 nm.

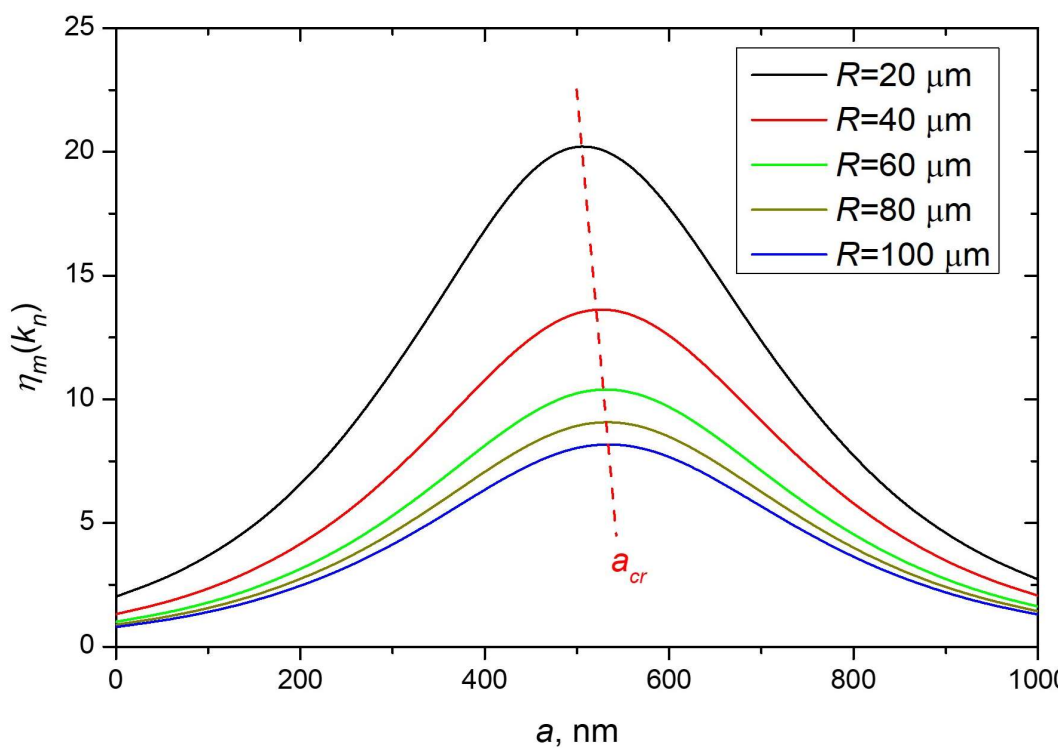

**Fig. 6.** Electric field enhancement $\eta_m(k_n)$ as a function of gap $a$ between the ring and the waveguide for different ring radii $R$.

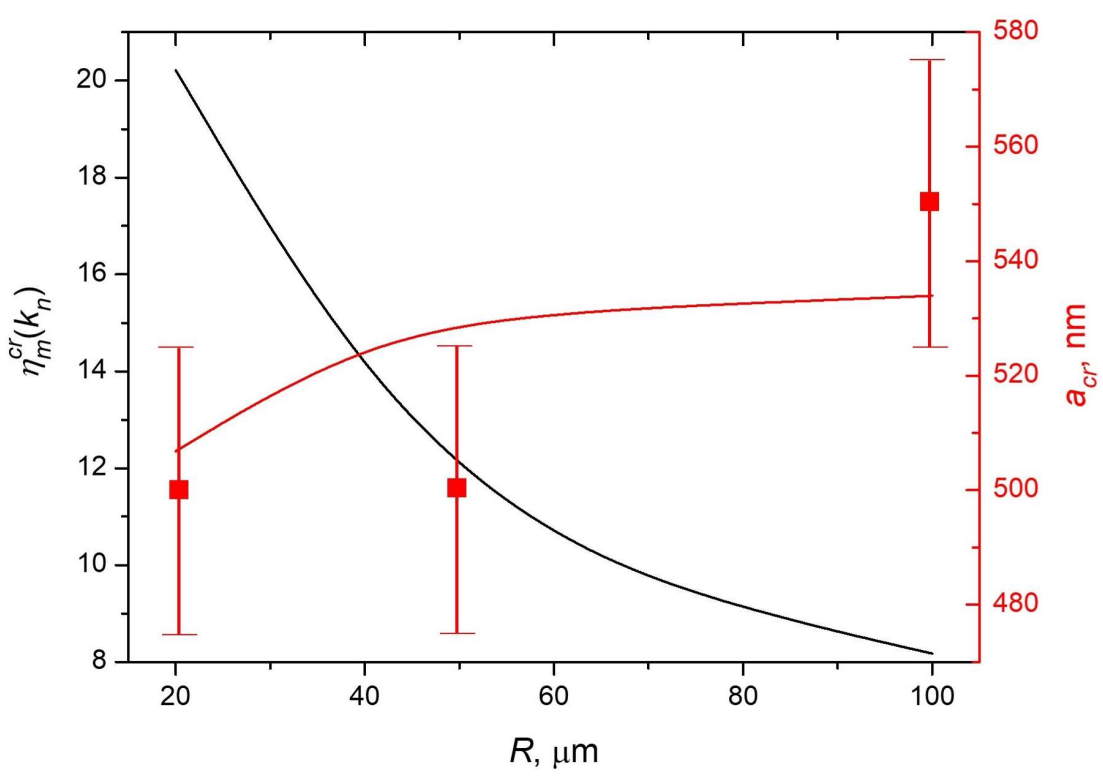

**Fig. 8.** Maximum field enhancement $\eta_m^{cr}(k_n)$ achieved at critical gap $a_{cr}$ as a function of $R$. The experimentally measured $a_{cr}$ are shown by the scatterers.

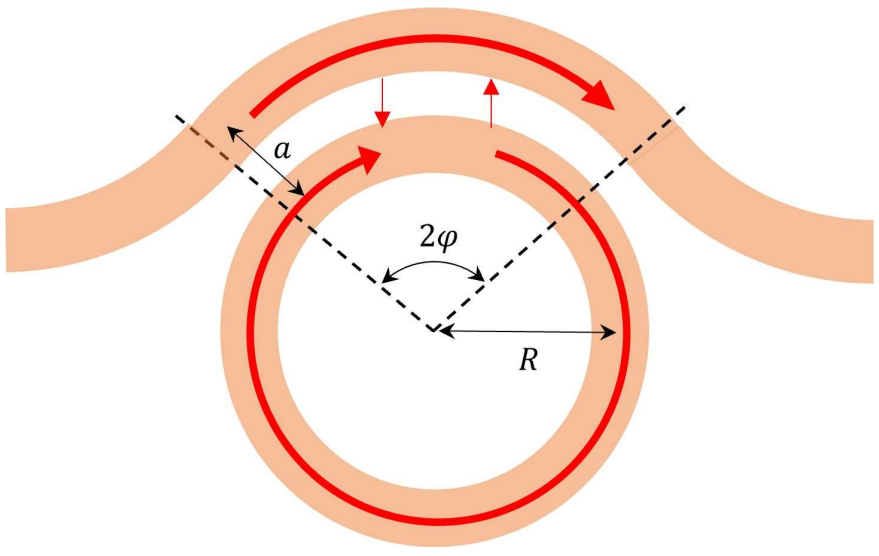

**Fig. 9.** Sketch of a pulley-type microring resonator.

The critical coupling realized at $a_{cr}$ results in $Q_m^{cr}(k_n) = \alpha(R)Q_m^r(k_n)$. Our model predicts that the coefficient $\alpha(R)$ approaches 1/2 for large rings with $\theta_{mm}^{out}(k_n,k) \approx \theta_{mm}^{in*}(k_n,k)$. For AlN microrings with $R$ in between of 20 and 100 µm, $\alpha(R)$ varies from 0.516 to 0.504. The weak dependence of $\alpha(R)$ for large rings, allows us to write

$$\eta_m^{cr}(k_n) \approx \frac{1}{2}Q_m^r(k_n)\left|\theta_{mm}^{in}(k_n,k_n)\right|. \quad (12)$$

This relation suggests that the main dependence of the maximum field enhancement $\eta_m^{cr}(k_n)$ on the coupler geometry comes from the in-coupling coefficient $\theta_{mm}^{in}(k_n,k_n)$. As $|\theta_{mm}^{in}(k_n,k_n)|$ continuously decreases with microring size, caused by weaker in-coupling into larger microring resonators, the maximum field enhancement $\eta_m^{cr}(k_n)$ drops with $R$, as shown in Fig. 8. At the same time, this trend can be changed by modifying the coupler geometry. For instance, the in-coupling coefficient of a pulley-type coupler shown in Fig. 9 is much less sensitive to $R$ and generally higher owing to a larger effective area of coupling. Its estimate (see Appendix D for details) gives

$$\theta_{mm}^{in}(k_n,k_n) = \theta_{mm}^{max}(k_n,k_n)\sin(k_n a\varphi), \quad (13)$$

where $\theta_{mm}^{max}(k_n,k_n)$ is the coupling coefficient amplitude. By changing the coupler's angular size $2\varphi$ from 0 to $\pi/(k_n a)$, we can continuously increase the in-coupling efficiency and, hence, the resonant field enhancement $\eta_m^{cr}(k_n)$ under critical coupling to the microring.

Our model for the first time gives the full-wave prediction for the in-coupling dependence of the ring field enhancement following the insights provided by the eigenmode decomposition method.

## V. CONCLUSION

In summary, we have developed a predictive model based on linearized Maxwell's equations that includes the leaky and confined modes of the waveguide and the microresonator into consideration. This is the first model enabling calculation of the coupling loss for microring resonators and providing full-wave insights into the physics underlying waveguide-ring coupling. The developed model has been verified using the quality factors measured in the linear regime of AlN resonators and has achieved an excellent agreement with the experimental results. To get insight into the linear field enhancement provided by the ring resonators, we have used the developed model to explore the regime of critical coupling crucial for nonlinear photonic applications, such as frequency comb generation. We have identified the critical gaps resulting in the highest linear field enhancement inside the microring and have demonstrated that the maximum field enhancement decreases with size of the microring resonator. The predicted critical gaps have been further confirmed with the experimentally measured extinction maxima for the bus waveguide transmission.

## APPENDIX A: STRUCTURAL DECOMPOSITION

Let us consider a system consisting of a coupled waveguide and a microring embedded into a material with the constant dielectric permittivity $\varepsilon_b$. We assume that the waveguide and the ring are made of the same material with the space-dependent permittivities $\varepsilon_{w,r}(\boldsymbol{r}) = \varepsilon_b + \delta\chi_{w,r}(\boldsymbol{r})$, where $\delta\chi_{w,r}(\boldsymbol{r})$ are their excessive susceptibilities. Linear excitation of electromagnetic fields in the waveguide and their further coupling in and out of the ring are fully described by linearized Maxwell's equations, which can be reduced to the following equation for the excited electric field $\boldsymbol{E}$ in the frequency domain $\omega$:

$$-\nabla\times\nabla\times\boldsymbol{E}(\boldsymbol{r},\omega) + \varepsilon_b\frac{\omega^2}{c^2}\boldsymbol{E}(\boldsymbol{r},\omega) =$$
$$-\frac{\omega^2}{c^2}[\delta\chi_w(\boldsymbol{r}) + \delta\chi_r(\boldsymbol{r})]\boldsymbol{E}(\boldsymbol{r},\omega) - \omega^2\mu_0\boldsymbol{P}_0(\boldsymbol{r},\omega), \quad (A1)$$

where $\boldsymbol{P}_0(\boldsymbol{r},\omega)$ is the polarization field of the external current pumping the waveguide, $c$ is the speed of light in vacuum, and $\mu_0$ is the magnetic constant.

Equation (A1) can be structurally decomposed into the two coupled problems of (i) the single waveguide and (ii) the single microring. For this, we put

$$\boldsymbol{E}(\boldsymbol{r},\omega) = \boldsymbol{E}_w(\boldsymbol{r},\omega) + \boldsymbol{E}_r(\boldsymbol{r},\omega), \quad (A2)$$

where $\boldsymbol{E}_w(\boldsymbol{r},\omega)$ is the field excited in the waveguide problem:

$$-\nabla\times\nabla\times\boldsymbol{E}_w(\boldsymbol{r},\omega)+\varepsilon_w(\boldsymbol{r})\frac{\omega^2}{c^2}\boldsymbol{E}_w(\boldsymbol{r},\omega)= -\frac{\omega^2}{c^2}\delta\chi_w(\boldsymbol{r})\boldsymbol{E}_r(\boldsymbol{r},\omega)-\omega^2\mu_0\boldsymbol{P}_0(\boldsymbol{r},\omega), \quad \text{(A3)}$$

and $\boldsymbol{E}_r(\boldsymbol{r},\omega)$ is the field excited in the microring problem:

$$-\nabla\times\nabla\times\boldsymbol{E}_r(\boldsymbol{r},\omega)+\varepsilon_r(\boldsymbol{r})\frac{\omega^2}{c^2}\boldsymbol{E}_r(\boldsymbol{r},\omega)= -\frac{\omega^2}{c^2}\delta\chi_r(\boldsymbol{r})\boldsymbol{E}_w(\boldsymbol{r},\omega). \quad \text{(A4)}$$

Highlight, the waveguide problem contains only $\varepsilon_w(\boldsymbol{r})$ and $\delta\chi_w(\boldsymbol{r})$, while the ring problem is given by $\varepsilon_r(\boldsymbol{r})$ and $\delta\chi_r(\boldsymbol{r})$, manifesting that only the waveguide or the ring are present in the formulated problems. The terms in the right-hand sides of the obtained equations describe pumping of the waveguide and coupling between the two problems.

## APPENDIX B: SOLUTION OF THE WAVEGUIDE PROBLEM

The waveguide problem can easily be solved if we assume that the waveguide is lossless, i.e., when $\varepsilon_w(\boldsymbol{r})\in\mathbb{R}$. In this case, waveguide fields can be expanded over eigenfields as shown in Eq. (2), where $\boldsymbol{E}_m^w(\boldsymbol{r},k)$ are continuous functions of real-valued wavenumber $k$ given by the eigenvalue problem:

$$\nabla\times\nabla\times\boldsymbol{E}_m^w(\boldsymbol{r},k)-\ \varepsilon_w(\boldsymbol{r})\frac{\omega_m^2(k)}{c^2}\boldsymbol{E}_m^w(\boldsymbol{r},k)=0. \quad \text{(B1)}$$

If the waveguide is aligned along the $x$ direction, then its eigenfields can be written as follows:

$$\boldsymbol{E}_m^w(\boldsymbol{r}-\boldsymbol{r}_w,k)=\boldsymbol{E}_m(\tilde{y},\tilde{z},k)e^{ik\tilde{x}}, \quad \text{(B2)}$$

where $(\tilde{x},\tilde{y},\tilde{z})$ are the cartesian coordinates of the waveguide system with the center at $\boldsymbol{r}=\boldsymbol{r}_w$ (see Fig. 2). The cross-sectional profiles $\boldsymbol{E}_m(\tilde{y},\tilde{z},k)$ and eigen-frequencies $\omega_m(k)$ are unique for every eigenwave index $m$. Thus, the waveguide fields given by Eq. (2) represent superposition of the guided modes with $k=k_m(\omega)$, where $k_m(\omega)$ is the inverse function to $\omega_m(k)$, and leaky modes with $k\neq k_m(\omega)$.

In our consideration, we intentionally limit the dielectric permittivity $\varepsilon_w(\boldsymbol{r})$ to reals. This results in real-valued $\omega_m(k)$ necessary for the eigenfields $\boldsymbol{E}_m^w(\boldsymbol{r},k)$ to be orthogonal in the Hilbert space and feature finite norms $\|\boldsymbol{E}_m^w(k)\|$:

$$\int\varepsilon_w(\boldsymbol{r})\,\boldsymbol{E}_n^{w*}(\boldsymbol{r},k)\cdot\boldsymbol{E}_m^w(\boldsymbol{r},k)\,d^3r= 2\pi\|\boldsymbol{E}_m^w(k)\|^2\delta_{nm}\delta(k-k'), \quad \text{(B3)}$$

$$\|\boldsymbol{E}_m^w(k)\|^2=\int\varepsilon_w(\tilde{y},\tilde{z})|\boldsymbol{E}_m(\tilde{y},\tilde{z},k)|^2\,d\tilde{y}\,d\tilde{z}. \quad \text{(B4)}$$

By applying eigenmode decomposition (2) for waveguide equation (A3) with the use of eigenfield orthogonality (B3), we get the equation

$$\left[\frac{\omega_m^2(k)}{\omega^2}-1\right]A_m^w(k,\omega)= \frac{\int\boldsymbol{E}_m^{w*}(\boldsymbol{r},k)\cdot\left[\delta\chi_w(\boldsymbol{r})\,\boldsymbol{E}_r(\boldsymbol{r},\omega)+\frac{\boldsymbol{P}_0(\boldsymbol{r},\omega)}{\varepsilon_0}\right]d^3r}{2\pi\|\boldsymbol{E}_m^w(k)\|^2} \quad \text{(B5)}$$

for evolution of the waveguide mode amplitudes in the frequency domain.

## APPENDIX C: SOLUTION OF THE RING PROBLEM

Solution of the ring problem is slightly more complicated compared to the waveguide one. Even if the ring material is lossless, eigenfrequencies of the ring are always complex due to their bending loss. As a result, ring eigenfields are not orthogonal in the Hilbert space and cannot be used for field decomposition. However, bending losses of ring eigenfields decrease with the ring radius $R$ (see Fig. 4 for total intrinsic loss rates). Thus, microring eigenfrequencies can be approximated with the real-valued ones for the rings sufficiently large compared to the excited wavelength. In this case, the ring eigenfields can be approximately written with waveguide eigenfields (B1). If the ring cross-sectional profile is the same as that of the waveguide, then the ring eigenfields can be written as follows:

$$\boldsymbol{E}_m^r(\boldsymbol{r}-\boldsymbol{r}_r,k_n)=\boldsymbol{E}_m(\tilde{r},\tilde{z},k_n)e^{ik_nR\tilde{\phi}}, \quad \text{(C1)}$$

where $(\tilde{r},\tilde{\phi},\tilde{z})$ are the cylindrical coordinates of the ring system with the center at $\boldsymbol{r}=\boldsymbol{r}_r$ (see Fig. 2). The ring field periodicity supports only the discrete azimuthal wavenumbers $k_n=n/R$. Finally, the ring fields can be expanded over large indices $|n|\gg1$ obeying the condition of small bending losses, as shown in Eq. (1). Note, most of the ring modes in this decomposition are leaky at a given frequency $\omega$. Only the modes that satisfy the condition $k_n=k_m(\omega)$ are guided and confine their energy inside the ring.

Finally, with the use of orthogonality of the ring eigenfield in the Hilber space:

$$\int\varepsilon_r(\boldsymbol{r})\,\boldsymbol{E}_n^{r*}(\boldsymbol{r},k_s)\cdot\boldsymbol{E}_m^r(\boldsymbol{r},k_p)\,d^3r= 2\pi R\|\boldsymbol{E}_m^r(k_s)\|^2\delta_{nm}\delta_{sp}, \quad \text{(C2)}$$

$$\|\boldsymbol{E}_m^r(k_s)\|^2=\int\varepsilon_r(\tilde{r},\tilde{z})|\boldsymbol{E}_m(\tilde{r},\tilde{z},k_s)|^2\,d\tilde{r}\,d\tilde{z}, \quad \text{(C3)}$$

we can get the equation

$$\left[\frac{\omega_m^2(k_n)}{\omega^2}-1\right]A_m^r(k_n,\omega)=\frac{\int\boldsymbol{E}_m^{r*}(\boldsymbol{r},k_n)\cdot\boldsymbol{E}_w(\boldsymbol{r},\omega)\delta\chi_r(\boldsymbol{r})d^3r}{2\pi R\|\boldsymbol{E}_m^r(k_n)\|^2} \quad \text{(C4)}$$

for the ring mode amplitudes $A_m^r(k_n,\omega)$ in the frequency domain.

## APPENDIX D: MASTER EQUATIONS

The right-hand sides of Eqs. (B5) and (C4) give us the contributions of pumping, in-coupling, and out-coupling. To separate those processes, we split the waveguide mode amplitudes: $A_m^w(\omega,k)=A_m^{w,in}(\omega,k)+A_m^{w,out}(\omega,k)$, where $A_m^{w,in}(\omega,k)$ is the waveguide field in-coupling to the ring, and $A_m^{w,out}(\omega,k)$ is the field out-coupling from the ring. Then, the in-coupling field is given by Eq. (3) with

$$A_m^0(\omega,k)=\frac{1}{2\pi\varepsilon_0}\int\frac{\boldsymbol{E}_m^{w*}(\boldsymbol{r},k)\cdot\boldsymbol{P}_0(\boldsymbol{r},\omega)}{\|\boldsymbol{E}_m^w(k)\|^2}d^3r, \quad \text{(D1)}$$

while the out-coupling fields obey the equation

$$\left[\frac{\omega_m^2(k)}{\omega^2}-1\right]A_m^{w,out}(\omega,k)= R\sum_{n=-\infty}^{\infty}\sum_{m'}\theta_{mm'}^{out}(k_n,k)A_{m'}^r(\omega,k_n) \quad \text{(D2)}$$

with the out-coupling coefficients

$$\theta^{out}_{mm'}(k_n,k) = \int \frac{\delta\chi_w(r)}{2\pi} \frac{\mathbf{E}^{w*}_m(r,k)\cdot \mathbf{E}^r_{m'}(r,k_n)}{\|\mathbf{E}^w_m(k)\|^2} d^3r. \quad \text{(D3)}$$

For the ring modes, we obtain the following equation:

$$\left[\frac{\omega_m^2(k_n)}{\omega^2} - 1\right] A^r_m(\omega,k_n) = \int_{-\infty}^{\infty} dk \sum_{m'} \theta^{in}_{mm'}(k_n,k) A^w_{m'}(\omega,k), \quad \text{(D4)}$$

where

$$\theta^{in}_{mm'}(k_n,k) = \int \frac{\delta\chi_r(r)}{2\pi R} \frac{\mathbf{E}^{r*}_m(r,k_n)\cdot \mathbf{E}^w_{m'}(r,k)}{\|\mathbf{E}^r_m(k_n)\|^2} d^3r \quad \text{(D5)}$$

are the in-coupling coefficients.

Note, the master equations were derived under the assumption of straight waveguide. However, they are also valid for a curved waveguide, if its bending loss is negligibly small, similar to those of the microring ones. In this case, the derived theory is readily applicable to pulley-type ring resonators shown in Fig. 9. These resonators provide an additional degree of control for the effective coupling by varying the coupler's angular size $2\varphi$ and feature generally higher coupling efficiency. Their out-coupling and in-coupling coefficients can be estimated as follows

$$\theta^{out}_{mm'}(k_n,k) = \frac{\sin([k(R+a)-k_nR]\varphi)}{k(R+a)-k_nR} \times \int \frac{\delta\chi_w(\tilde{r},\tilde{z})}{\pi} \frac{\mathbf{E}^{w*}_m(\tilde{r}-a,\tilde{z},k)\cdot \mathbf{E}_{m'}(\tilde{r},\tilde{z},k_n)}{\|\mathbf{E}^w_m(k_n)\|^2} d\tilde{r}\, d\tilde{z}, \quad \text{(D6)}$$

$$\theta^{in}_{mm'}(k_n,k) = \frac{\sin([k(R+a)-k_nR]\varphi)}{k(R+a)-k_nR} \times \int \frac{\delta\chi_r(\tilde{r},\tilde{z})}{\pi} \frac{\mathbf{E}^{r*}_m(\tilde{r},\tilde{z},k_n)\cdot \mathbf{E}_{m'}(\tilde{r}-a,\tilde{z},k)}{\|\mathbf{E}^r_m(k_n)\|^2} d\tilde{r}\, d\tilde{z}, \quad \text{(D7)}$$

where the waveguide fields are written in the cylindrical system of the microring. The above estimates are obtained after integration over the angle $\phi$ from $-\varphi$ to $\varphi$ for the coupler section (see Fig. 9).

## APPENDIX E: COUPLING LOSS RATES

To get the ring mode coupling loss rates, we use Eqs. (D2) and (D4). Together, they result in the following evolution equation for the ring modes:

$$\left[\frac{\omega_m^2(k_n)}{\omega^2} - 1 - \beta_m(\omega,k_n,k_n)\right] A^r_m(\omega,k_n) = B^{in}_m(\omega,k_n) + \sum_{n'\neq n} \beta_m(\omega,k_n,k_{n'}) A^r_m(\omega,k_{n'}) \quad \text{(E1)}$$

with

$$B^{in}_m(\omega,k_n) = \int_{-\infty}^{\infty} dk \sum_{m'} \theta^{in}_{mm'}(k_n,k) A^{w,in}_{m'}(\omega,k), \quad \text{(E2)}$$

$$\beta_m(\omega,k_n,k_{n'}) = R\int_{-\infty}^{\infty} dk \sum_{m'} \frac{\omega^2 \theta^{in}_{mm'}(k_n,k)}{\omega^2_{m'}(k)-\omega^2} \sum_{m''} \theta^{out}_{m'm''}(k_{n'},k). \quad \text{(E3)}$$

The imaginary part of $\beta_m(\omega,k_n,k_n)$ in the left-hand side of Eq. (E1) defines the ring modes coupling loss rates

$$\gamma^c_m(\omega,k_n) = \frac{1}{2}\mathrm{Im}[\beta_m(\omega,k_n,k_n)] = \frac{\pi R}{4}\, \mathrm{Re} \sum_{m'} \left|\frac{\omega}{v^g_{m'}(k_{m'}(\omega))}\right| \theta^{in}_{mm'}(k_n,k_{m'}(\omega)) \times \sum_{m''} \theta^{out}_{m'm''}(k_n,k_{m'}(\omega)). \quad \text{(E4)}$$

Note that in a multiple-eigenwave case, the coupling loss rate of a particular ring mode with index $m$ depends on characteristics of all $m'$ eigenwaves supported by the ring and the waveguide. However, if the coupling is mainly driven by the $m-m$ type of interaction, with the leading $\theta^{in}_{mm}$ and $\theta^{out}_{mm}$ coefficients, then Eq. (E4) reduces to Eq. (7) for a single eigenwave operation regime.

## APPENDIX F: WAVEGUIDE TRANSMISSION

According to Eqs. (3) and (D2), ring mode amplitudes define the waveguide fields:

$$\frac{A^w_m(\omega,k)}{A^{w,in}_m(\omega,k)} = 1 + R\sum_{n=-\infty}^{\infty} \sum_{m'} \frac{\theta^{out}_{mm'}(k_n,k) A^r_{m'}(\omega,k_n)}{A^0_m(\omega,k)}. \quad \text{(F1)}$$

If $Q$-factors of the ring resonances are sufficiently high and the microring spectral lines with different eigenwave indices $m$ do not overlap, then in the vicinity of the resonant angular frequency $\omega^{res}_m(k_n)$ Eq. (F1) can be reduced to the following relation:

$$\frac{A^w_m(\omega,k)}{A^{w,in}_m(\omega,k)} \approx 1 + R\frac{\theta^{out}_{mm}(k_n,k) A^r_m(\omega,k_n)}{A^0_m(\omega,k)}. \quad \text{(F2)}$$

Under critical coupling to the resonant ring mode, when the waveguide transmission drops to zero, the waveguide mode amplitude is $A^w_m(\omega^{res}_m(k_n),k_n) = 0$. Finally, this gives us the following relation for the critically coupled ring amplitude $A^{r,cr}_m(k_n)$:

$$A^r_m(\omega^{res}_m(k_n),k_n) \approx A^{r,cr}_m(k_n) = -\frac{A^0_m(\omega^{res}_m(k_n),k_n)}{R\,\theta^{out}_{mm}(k_n,k_n)}. \quad \text{(F3)}$$

With introduction of $A^{r,cr}_m(k_n)$, we can rewrite the waveguide fields ratio as follows

$$\frac{A^w_m(\omega,k)}{A^{w,in}_m(\omega,k)} \approx 1 - \frac{A^r_m(\omega,k_n)}{A^{r,cr}_m(k_n)}. \quad \text{(F4)}$$

This ratio demonstrates the direct link between a minimum in the waveguide transmittance given by the left-hand side of Eq. (F4) and a maximum of the ring fields, allowing us to use any of them for $Q$-factor characterization.

## APPENDIX G: FABRICATION

AlN microring resonators are fabricated on 8-inch silicon (Si) wafer. The wafer is first deposited with 2.4 µm thick silicon dioxide ($SiO_2$) as the bottom cladding layer. AlN of thickness ~500 nm is then deposited on top of the bottom cladding layer using physical vapor deposition (PVD) method. This layer of AlN is subsequently subjected to high temperature annealing to improve its quality. A thin layer of $SiO_2$ (~200 nm thickness) is then deposited on AlN to act as a hard mask for AlN for subsequent process steps. Photolithography and etching are then conducted to imprint the resonators patterns on photoresist and transfer these patterns down to the $SiO_2$ hard mask and AlN device layer via etching. After the AlN resonators patterns are formed, a layer of $SiO_2$ of ~ 1.5 µm is blanket deposited on the AlN devices as a final top cladding layer. More details of the fabrication can be found in [25-27].

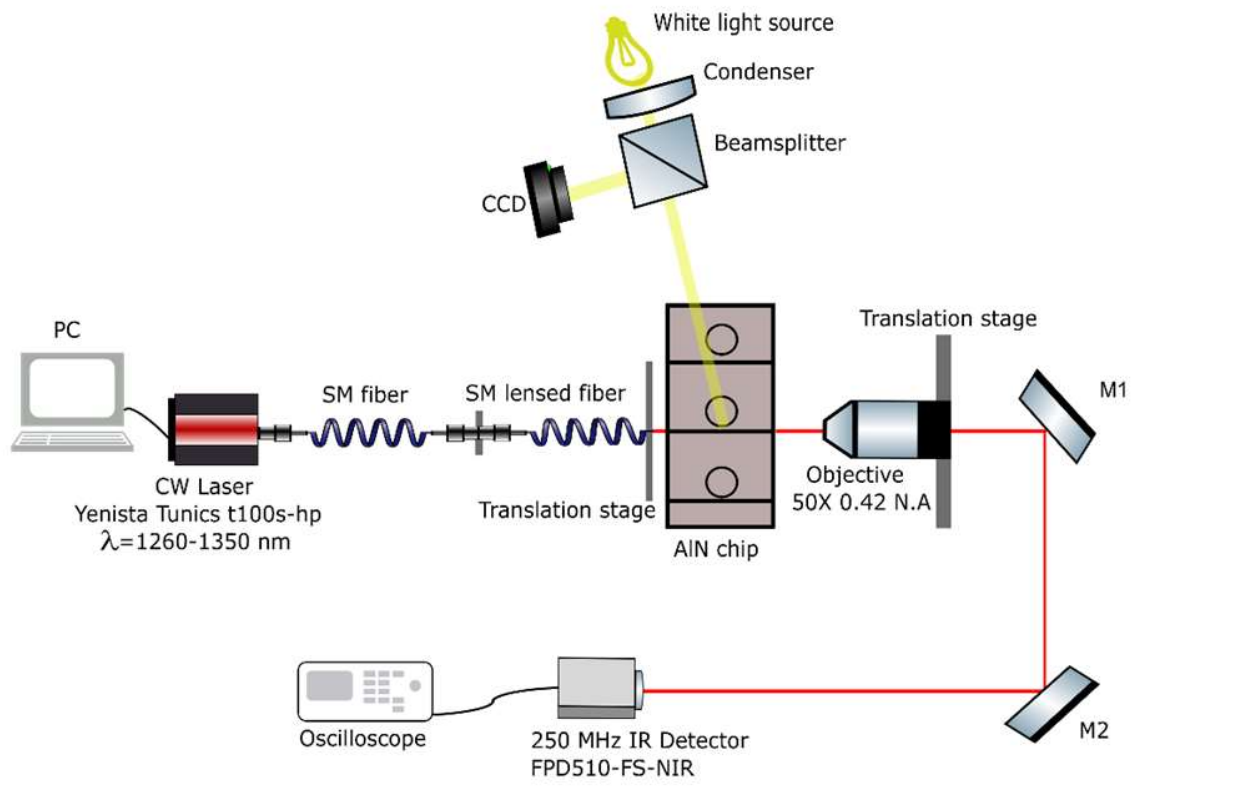

**Fig. H1.** Experimental setup used for transmission measurement.

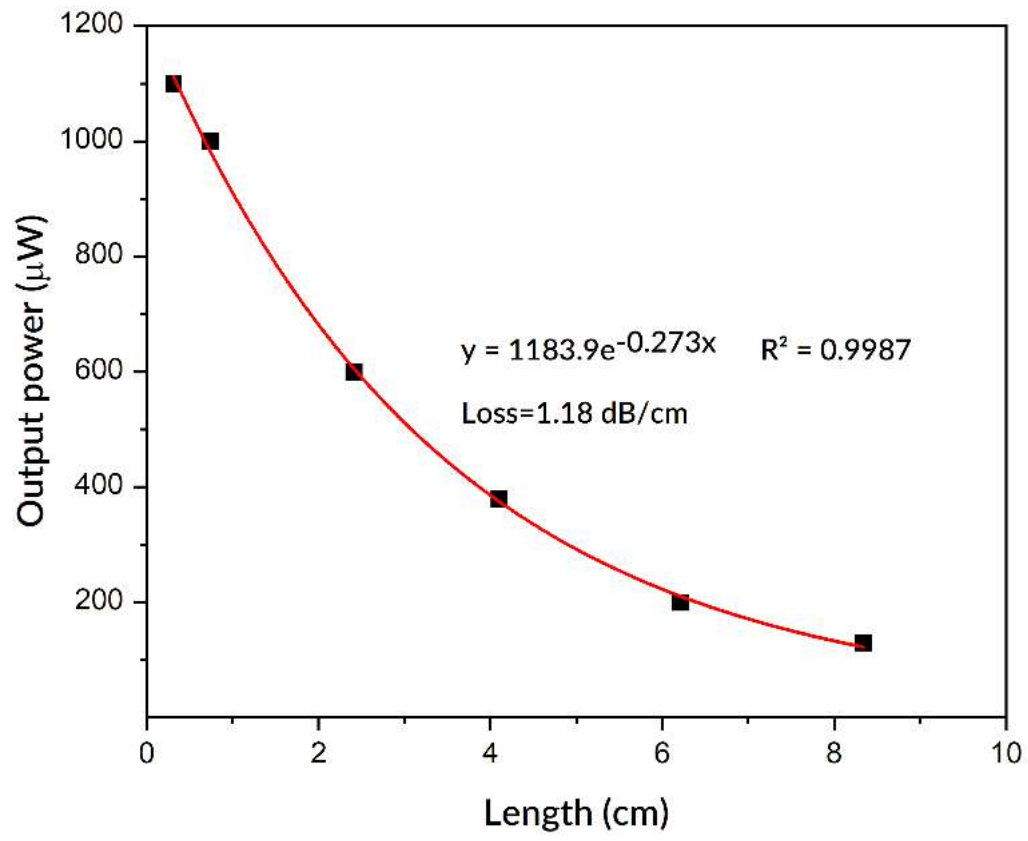

**Fig. I1.** Experimental output coupled power as a function of waveguide length.

## APPENDIX H: TRANSMISSION MEASUREMENTS

The $Q$-factor measurements are done for the waveguide transmittance using a Yenista Tunics 100S-HP tunable laser around the wavelength of 1.3 µm. The laser is edge coupled to the waveguide using a lensed fiber. The polarization of the laser is fixed to TE to ensure only a single transverse mode is excited inside the waveguide. Brightfield imaging system is used to align the fiber tip to the waveguide. The output from the fiber is collected using an objective lens (50X, 0.42N.A) on to a high-speed IR detector (FPD510-FS-NIR, 250MHz). The laser is scanned at the speed of 100nm/s across multiple resonances and then transmission of the ring-waveguide system is observed on an oscilloscope. The input power is controlled to ensure additional thermal effects, which can result in the broadening of the resonances, are avoided. The schematic for the experimental setup is shown in Fig. H1.

## APPENDIX I: INTRINSIC LOSS ESTIMATION

By using the cut-back method, experimental propagation loss of the fundamental mode at the wavelength of 1.3 µm excited in the waveguide was estimated to be 1.18 dB/cm, as shown in Fig. I1. This estimation gave us the waveguide intrinsic loss rate $\gamma_m^w$ of $1.3 * 10^{-6}$.

The experimentally estimated rate $\gamma_m^w$ was reproduced in COMSOL Multiphysics eigenmode solver for a straight waveguide by introducing finite dissipation to the materials used. Those material data were then input in the same eigenmode solver for standing alone ring resonators modeling the intrinsic loss rates $\gamma_m^r$ that account for both propagation and bending losses. Calculation of the rates $\gamma_m^w$ and $\gamma_m^r$ was done in 2D eigenmode solver. In both the cases, the number of mesh elements over the wavelength was >13 for the external area and >26 for the internal part of the waveguide/ring (compared to the minimum requirement of 8 mesh elements per wavelength) that guaranteed high accuracy for the derived loss rates. The obtained microring loss rates appear weakly dependent of ring radius $R$, as shown in Fig. 4.

## ACKNOWLEDGMENTS

The authors acknowledge A*STAR strategic program number C210917001.